\documentclass[%
reprint,
nofootinbib,
 amsmath,
 amssymb,
 aps,
 pra,
floatfix,
]{revtex4-1}

\usepackage{graphicx}
\usepackage{dcolumn}
\usepackage{bm}

\usepackage[utf8]{inputenc}
\usepackage{mathtools}
\usepackage{amsfonts}
\usepackage{mathalfa}
\usepackage{mathrsfs}
\usepackage[english]{babel}
\usepackage{csquotes}

\usepackage{hyperref}
\hypersetup{colorlinks=true, urlcolor=blue, linkcolor=blue, citecolor=blue}

\begin{document}

\title{Quantum Algorithm for the Vlasov Equation}

\author{Alexander Engel}
 \email{alen3220@colorado.edu}
\author{Graeme Smith}
 \altaffiliation[Also at ]{JILA, University of Colorado, Boulder, Colorado 80309, USA}
\author{Scott E. Parker}
\affiliation{%
 Department of Physics, University of Colorado, Boulder, Colorado 80309, USA}%
 
\date{\today}

\begin{abstract}

The Vlasov-Maxwell system of equations, which describes classical plasma physics, is extremely challenging to solve, even by numerical simulation on powerful computers.  By linearizing and assuming a Maxwellian background distribution function, we convert the Vlasov-Maxwell system into a Hamiltonian simulation problem. Then for the limiting case of electrostatic Landau damping, we design and verify a quantum algorithm, appropriate for a future error-corrected universal quantum computer. While the classical simulation has costs that scale as $\mathcal{O}(N_v t)$ for a velocity grid with $N_v$ grid points and simulation time $t$, our quantum algorithm scales as $\mathcal{O}(\text{polylog}(N_v) t/\delta)$ where $\delta$ is the measurement error, and weaker scalings have been dropped. Extensions, including electromagnetics and higher dimensions, are discussed. A quantum computer could efficiently handle a high-resolution, six-dimensional phase-space grid, but the $1/\delta$ cost factor to extract an accurate result remains a difficulty. This paper provides insight into the possibility of someday achieving efficient plasma simulation on a quantum computer.

\end{abstract}

\maketitle

\section{Introduction}

Quantum computers show enormous promise for solving classes of problems for which a quantum algorithm can obtain an exponential speedup over the classical counterpart. Naturally, this includes the simulation of quantum systems \cite{Feynman1982, Babbush18}, but it also includes problems that are not inherently quantum, such as integer factorization \cite{Shor1997} and solving linear systems \cite{Harrow2009, QLSA2017}. However, it remains unclear how many classical computational tasks requiring the largest supercomputers\cite{Kothe07} could be sped up using a future error-corrected quantum computer.

One such computationally extreme example is the kinetic plasma problem\cite{Lapenta2015, Howard2015, Ku2018}. High-temperature plasmas occur in a wide range of applications, including fusion, space, solar, and astrophysical contexts. Such plasmas are more complex than conventional fluids because near-range forces are Debye shielded\cite{Debye1923, Nicholson1983}.  This leads to dynamics far from thermal equilibrium, requiring the time evolution of a smooth, six-dimensional distribution function $f({\bf x}, {\bf v}, t)$, describing the phase-space density of particles. 

High-temperature plasmas are fundamentally described by the Vlasov-Maxwell system of equations which arise from Newton's second law and Maxwell's equations, neglecting short-range Coulomb interactions.  The Vlasov equation\cite{Vlasov1938, Nicholson1983} is an excellent testbed for applying quantum algorithms to classical computational mathematics because it is a hyperbolic partial differential equation describing a conservation law (conservation of particles) that has features similar to equations in computational fluid dynamics and other applications\cite{Leveque92}.

We study the Vlasov-Maxwell system and find that, when linearized about a Maxwellian equilibrium, the time evolution can be expressed in a unitary fashion. This is then mathematically equivalent to the evolution of a quantum system. On a universal quantum computer, the time evolution of arbitrary quantum systems is achieved with Hamiltonian simulation algorithms. These algorithms have undergone significant development in recent years\cite{Berry2015, Berry2016, Berry2017}, culminating in the achievement of optimal query complexity for sparse Hamiltonians \cite{Low2017}. That may be enough for the simulation of physical quantum systems, but for more general computations the efficient handling of non-sparse Hamiltonians can be important. On this front there has also been progress: a number of algorithms which do not rely on sparsity have been developed\cite{Childs2010, Qubitization, Wang2018}. However, efficient simulation is not possible for all non-sparse Hamiltonians\cite{NonsparseLimits}.

The efficiency of a quantum algorithm that time-evolves the linearized Vlasov-Maxwell system is unclear since the associated Hamiltonian is not sparse. Additionally, encoding initial data into a quantum state and extracting final data out can be expensive. To explore these issues, we fully detail an algorithm for a simple limiting form of the full system. This will correspond to the electrostatic Landau damping problem, which is an important kinetic problem in plasma physics. Although the simulation of Landau damping is not computationally demanding, we show that the quantum algorithm scales differently than its classical counterpart. That could translate into a large speedup in generalizations of the algorithm, which are discussed. This paper is intended as a first step in exploring the potential of quantum computation applied to plasma physics problems.

An outline of the paper follows. In Sec.~\ref{sec:formulation} we analyze the linearized Vlasov-Maxwell system. We then focus on the electrostatic Landau damping problem.  In Sec.~\ref{sec:Hsim} we detail the Hamiltonian simulation algorithm used to solve the Landau damping problem.  In Sec.~\ref{sec:tests} the results of numerical tests (using a classical computer) are given, and in Sec.~\ref{sec:measure} we discuss initialization and measurement. Section \ref{sec:ext} explores possible extensions of the current algorithm, and finally we summarize our findings in Sec.~\ref{sec:sum}.

\section{The Vlasov-Maxwell System}\label{sec:formulation}

We begin with the nonrelativistic Vlasov equation:
\begin{equation}\label{eq:fullVlasov}
\frac{\partial f_s}{\partial t} + {\bf v} \cdot \boldsymbol{\nabla} f_s +
\frac{q}{m}\left( {\bf E} + {\bf v} \times {\bf B} \right) \cdot \frac{\partial f_s}{\partial {\bf v}} = 0,
\end{equation}

\noindent where $f_s = f_s({\bf x},{\bf v}, t)$ is the distribution function for species $s$, i.e. electrons or a type of ions. We solve for the behavior of small-amplitude waves by linearizing Eq.~(\ref{eq:fullVlasov}) and assuming a Maxwellian background distribution $f_M({\bf v})$, at rest. The zeroth-order equation is satisfied only if the background electric field ${\bf E}_0$ vanishes. The first-order equation for electrons is

\begin{equation}\label{eq:partVlasov}
\frac{\partial f_1}{\partial t} + {\bf v} \cdot \boldsymbol{\nabla} f_1 =
\frac{e}{m} \left(-{\bf E}_1 \cdot {\bf v} f_M + ({\bf v} \times {\bf B}_0) \cdot \frac{\partial f_1}{\partial {\bf v}}\right).
\end{equation}

Assuming stationary ions, the time evolution is governed by Eq.~(\ref{eq:partVlasov}) and Maxwell's Equations. To adapt these to a quantum algorithm we apply a few transformations. We Fourier-transform in space, introducing the wavevector ${\bf k}$, and switch to these dimensionless variables:
\begin{align*}
\hat{\bf k} &= \lambda_{De} {\bf k}, & \hat{t} &= \omega_{pe} t, \\
\hat{\bf v} &= \frac{\bf v}{\lambda_{De} \omega_{pe}}, & \hat{f} &= \frac{(\lambda_{De} \omega_{pe})^3}{n_e} f, \\
\hat{\bf E} &= \frac{e \lambda_{De}}{k_B T_e} {\bf E}, & \hat{\bf B} &= \frac{e \lambda_{De}}{k_B T_e} c {\bf B},
\end{align*}
where $T_e$ is the electron temperature, $\omega_{pe}$ is the electron plasma frequency, $n_e$ is the electron number density, and $\lambda_{De}$ is the Debye length with ions neglected. Applying these changes, Eq.~(\ref{eq:partVlasov}) becomes

\begin{equation}\label{eq:Vlasov2}
\frac{\partial \tilde{f}}{\partial t} = -i {\bf k} \cdot {\bf v} \tilde{f} - \tilde{\bf E} \cdot {\bf v} f_M + \frac{1}{c_n} ({\bf v} \times {\bf B}_0) \cdot \frac{\partial \tilde{f}}{\partial {\bf v}},
\end{equation}
where all quantities are now the dimensionless versions, $\sim$ identifies variables that are Fourier components for wavevector ${\bf k}$, the $1$ subscripts have been dropped, and $c_n := \frac{c}{\lambda_{De} \omega_{pe}}$ is the speed of light expressed in units of $\lambda_{De} \omega_{pe}$. Applying the same transformations to Maxwell's Equations gives

\begin{equation}\label{eq:Amp}
\frac{\partial \tilde{\bf E}}{\partial t} = i c_n {\bf k} \times \tilde{\bf B} + \int {\bf v} \tilde{f} d^3 v,
\end{equation}
\begin{equation}\label{eq:Faraday}
\frac{\partial \tilde{\bf B}}{\partial t} = -i c_n {\bf k} \times \tilde{\bf E}.
\end{equation}

Equations (\ref{eq:Vlasov2}), (\ref{eq:Amp}), and (\ref{eq:Faraday}) are sufficient to time-evolve the variables $\tilde{\bf E}$, $\tilde{\bf B}$, and $\tilde{f}({\bf v})$. To employ Hamiltonian simulation algorithms, we also need the system's time evolution to be unitary. Consider the following change of variables:\footnote{An arbitrary $e^{i \phi({\bf v})}$ factor can be included in Eq.~(\ref{eq:rescale}) without breaking the unitarity of the time evolution.}
\begin{equation}\label{eq:rescale}
\tilde{f}'({\bf v}) := i \sqrt{\frac{\xi}{f_M({\bf v})}} \tilde{f}({\bf v}),
\end{equation}
and the real constant $\xi$ will be chosen later. This turns Eqs. (\ref{eq:Vlasov2}) and (\ref{eq:Amp}) into

\begin{equation}\label{eq:Vlasov3}
\frac{\partial \tilde{f}^{\prime}}{\partial t} = -i {\bf k} \cdot {\bf v} \tilde{f}^{\prime} - i \mu \tilde{\bf E} \cdot {\bf v} + \frac{1}{c_n} ({\bf v} \times {\bf B}_0) \cdot \frac{\partial \tilde{f}'}{\partial {\bf v}},
\end{equation}

\begin{equation}\label{eq:Amp2}
\frac{\partial \tilde{\bf E}}{\partial t} = i c_n {\bf k} \times \tilde{\bf B} - i \int {\bf v} \mu \tilde{f}^{\prime} \frac{d^3 v}{\xi},
\end{equation}
where $\mu({\bf v}) := \sqrt{\xi f_M({\bf v})}$. The time evolution is unitary if it can be written as $d \vert x \rangle/dt = -i\hat{H}_{\text{eff}} \vert x \rangle$ where $\hat{H}_{\text{eff}} = \hat{H}_{\text{eff}}^\dagger$ is the effective Hamiltonian of the system, and $\vert x \rangle$ is a quantum state with the variables encoded as amplitudes. Now, we consider the terms in Eqs. (\ref{eq:Faraday}), (\ref{eq:Vlasov3}), and (\ref{eq:Amp2}). First, the evolution within the velocity space is generated by $\hat{H}_v$ with
\begin{equation}\label{eq:velop}
\hat{H}_v = {\bf k} \cdot {\bf v} + \frac{i}{c_n} ({\bf v} \times {\bf B}_0) \cdot \frac{\partial}{\partial {\bf v}}.
\end{equation}
This is Hermitian since
\begin{equation}
\int g({\bf v}) \hat{H}_v h({\bf v}) d^3v = \int h({\bf v}) \hat{H}^*_v g({\bf v}) d^3v
\end{equation} for normalizable functions $g({\bf v})$ and $h({\bf v})$, using integration by parts. The behavior of the last term in Eq.~(\ref{eq:velop}) is further explained by Eq.~(\ref{eq:Lv}). Next, defining ${\bf k} := (0, 0, k)$, the electromagnetic (EM) part of the evolution can be expressed as
\begin{equation}\label{eq:EM}
\frac{\partial \vert EM \rangle}{\partial t} =
-i\hat{H}_{l} \vert EM \rangle = -i c_n k
\begin{bmatrix}
0 & 0 & 0 & 1 \\
0 & 0 & -1 & 0 \\
0 & -1 & 0 & 0 \\
1 & 0 & 0 & 0
\end{bmatrix}
\begin{bmatrix}
E_x \\
E_y \\
B_x \\
B_y
\end{bmatrix}.
\end{equation}
Clearly, $\hat{H}_{l}$ is a Hermitian matrix. Lastly, there is the coupling between the velocity distribution and the electric field. Here the Hermitian restriction can be expressed in terms of coupling constants $C_p ({\bf v})$:
\begin{align}\label{eq:coupling}
\frac{\partial \tilde{f}^{\prime}({\bf v})}{\partial t} &=  -i C_p({\bf v}) E_p, & \frac{\partial E_p}{\partial t} &= -i C^*_p({\bf v}) \tilde{f}^{\prime}({\bf v}),
\end{align}
where $E_p$ is the $p^{th}$ component of ${\bf E}$. From Eq.~(\ref{eq:Vlasov3}) we find $C_p({\bf v}) = \mu v_p$ while from Eq.~(\ref{eq:Amp2}) we get $C_p({\bf v}) = \mu v_p d^3 v/\xi$. Consequently we need to choose $\xi = d^3v$. In practice we will have a uniform, finite grid in velocity space in which case $\xi$ will be the volume of a single velocity grid cell. Now, we can formally consider the coupling constants to be components of an operator $\hat{H}_c$, which acts on a space that is the direct sum of the spaces in which $\tilde{f}^{\prime}({\bf v})$ and ${\bf E}$ are specified. Then Eq. (\ref{eq:coupling}) says that $\hat{H}_c$ is Hermitian, so $\hat{H}_{\text{eff}} = \hat{H}_v + \hat{H}_l + \hat{H}_c$ is Hermitian, and thus the evolution generated by $\hat{H}_{\text{eff}}$ is unitary.

\subsection{The Landau Damping Problem}\label{sec:Landau}

Now that we have unitary time evolution, in principle Hamiltonian simulation can be applied to time-evolve Eqs. (\ref{eq:Faraday}), (\ref{eq:Vlasov3}), and (\ref{eq:Amp2}), but designing an efficient quantum algorithm is still a challenging task. We will demonstrate an algorithm for the electrostatic case with ${\bf B}_0 = {\bf 0}$. This still has the coupling term [Eq.~(\ref{eq:coupling})], which is tricky to handle efficiently, and it captures the physics of Landau damping\cite{Landau1946}. In developing the quantum algorithm we target and achieve costs scaling only logarithmically with the grid size, in contrast to the usual linear scaling of classical algorithms. Although a large grid will not be necessary in our simulation of Landau damping, extensions of the algorithm to handle problems that involve very large grids are conceivable. These are discussed in Sec.~\ref{sec:ext}.

In the electrostatic case there is only an electric field along ${\bf k}$, i.e. only $E_z$ is nonzero. Moreover, the electric field evolution can be determined using only a 1D velocity space: the velocity space, represented by a uniform 3D grid indexed by ${\bf j}$, can be reduced to 1D by introducing
\begin{align}
\tilde{F}_{j_z} &:= \sum_{j_x, j_y} \tilde{f}_{\bf j} \Delta^2 v, & G_{j_z} &:= \sum_{j_x, j_y} f_M({\bf v_j}) \Delta^2 v,
\end{align}
\begin{equation}
\tilde{F}'_{j_z} := i \sqrt{\frac{\Delta v}{G_{j_z}}} \tilde{F}_{j_z},
\end{equation}
in terms of which Eqs. (\ref{eq:Vlasov2}) and (\ref{eq:Amp}), now with ${\bf B}_0 = {\bf 0}$, become

\begin{equation}\label{eq:fprime}
\frac{\partial \tilde{F}^\prime_j}{\partial t} = -i k v_j \tilde{F}'_j - i \alpha_j \tilde{E} v_j,
\end{equation}

\begin{equation}
\frac{\partial \tilde{E}}{\partial t} = - i \sum_{j} \alpha_j \tilde{F}'_j v_j,
\end{equation}
where $\alpha_j := \sqrt{\Delta v G_j}$, and all quantities that were previously vectors are now their $z$ components, e.g. $j := {\bf j}_z$. To apply Hamiltonian simulation, we encode the data as the amplitudes of a quantum state, written in bra-ket notation as
\begin{equation}\label{eq:state}
\vert x\rangle = \eta \left(\sum_{j=0}^{N_v-1} \tilde{F}'_{j} \vert j \rangle + \tilde{E} \vert N_v\rangle \right),
\end{equation}
\noindent where $N_v$ is the total number of velocity grid cells, and $\eta := \left(\vert \tilde{E} \vert^2 + \sum_{j=0}^{N_v-1} \vert\tilde{F}'_{j}\vert^2\right)^{-1/2}$ normalizes the state. This state evolves via $d\vert x\rangle/dt = -i \hat{H} \vert x\rangle$ with a Hamiltonian given by
\begin{equation}\label{eq:Ham}
\hat{H} = \sum_{j=0}^{N_v-1} v_j \left[ k \vert j \rangle \langle j \vert + \alpha_j \left(\vert j \rangle \langle N_v \vert + \vert N_v \rangle \langle j \vert \right) \right].
\end{equation}

\section{Hamiltonian Simulation}\label{sec:Hsim}

Efficient Hamiltonian simulation of Eq.~(\ref{eq:Ham}) is somewhat difficult due to the matrix being non-sparse in the sense often assumed for quantum algorithms. Specifically, many Hamiltonian simulation techniques (e.g. \cite{Berry2007, Berry2016, Berry2017}) have costs that scale polynomially with the maximum number of nonzero entries in any row. Our Hamiltonian has $N_v$ nonzero entries in one row due to how the electric field couples with all velocities.

To avoid a $\text{poly}(N_v)$ cost scaling, we apply the Hamiltonian simulation technique developed by \textcite{Low2017} (detailed further in \cite{Qubitization, LowSpectral, PhaseVectorCalc}). This technique allows the Hamiltonian to be encoded in a general way, which creates possibilities for efficiently handling non-sparse cases. It also has optimal query complexity, scaling linearly with time plus a term logarithmic in the error. Still, we must show how the individual queries can be performed efficiently. In what follows, we detail our Hamiltonian simulation implementation and analyze its efficiency.

Our algorithm operates on a quantum state with $n_v + 7$ qubits where $n_v := \log_2(N_v)$. We divide this into five registers. Three of these, labeled by $b$, $q$, and $r$, are single-qubit registers, $a$ has four qubits, and $v$ has $n_v$ qubits. The main data are stored in the $r$ and $v$ registers as
\begin{equation}\label{eq:init}
\vert x \rangle_s = \eta \left(\sum_{j=0}^{N_v-1} \tilde{F}_{j}^\prime \vert 0 \rangle_r \vert j \rangle_v + \tilde{E} \vert 1 \rangle_r \vert 0 \rangle_v \right),
\end{equation}
where $s$ is used to denote the combined $r$ and $v$ registers. It is on this $s$ register that the Hamiltonian [Eq.~(\ref{eq:Ham})] acts, and the $s$ state components with indices in $(N_v, 2N_v)$ are unused. The other registers have ancilla qubits. They are initialized as $\vert 0 \rangle$, and at the end of the algorithm if any of them are measured to be nonzero then the algorithm has failed and must be rerun.

\begin{figure}[htbp]
\includegraphics[width=8.6cm]{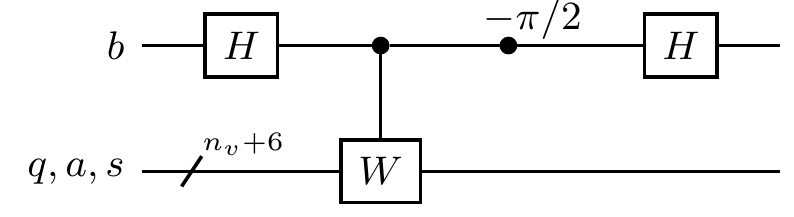}
\caption{Implementation of $V$. Gate $W$ is given by Fig.~\ref{cir:W}. The notation of an angle $\phi$ next to a filled dot represents the phase shift gate, i.e. the phase $\phi$ is applied controlled on that qubit. Our diagrams were created using the Quantikz package\cite{Quantikz}.}\label{cir:V}
\end{figure}

We describe the algorithm from the top down. Only the query implementation $U$ at the bottom is of our design. The higher operations are not specific to our Hamiltonian and are described by \textcite{Qubitization}, but we include them for completeness. Hamiltonian simulation is performed by the circuit
\begin{equation}\label{eq:Hcir}
\hat{C} := H_{bq}\left(\overset{\curvearrowleft}{\prod^{L/2}_{z = 1}} B_b(\theta_z)V^\dagger B_b(\phi_z) V B_b(\vartheta_z)\right)H_{bq},
\end{equation}
where $H_{bq}$ denotes a Hadamard gate, $\frac{1}{\sqrt{2}}\begin{psmallmatrix}1 & 1 \\ 1 & -1 \end{psmallmatrix}$, applied to both the $b$ and $q$ qubits; $B_b(\Phi)$ is the phase shift gate, $\begin{psmallmatrix}1 & 0 \\ 0 & e^{i\Phi} \end{psmallmatrix}$, applied to the $b$ qubit; and $V$ is the quantum circuit depicted in Fig.~\ref{cir:V}. The ${}^{\curvearrowleft}$ over the product means that it is expanded in right-to-left fashion, e.g. the $z=1$ term is the first circuit operation. The angles used in the phase shift gates are determined\footnote{$\theta_{j+1} := \pi+\varphi_{2j+1}$, $\vartheta_{j+1} := -\varphi_{2j}$, $\phi_j := -\theta_j-\vartheta_j$} by the phase vector $\vec{\varphi}$ which is computed as explained by \textcite{PhaseVectorCalc}. $\vec{\varphi}$ depends only on the simulation time and error threshold, and its length $L$ is proportional to the query complexity. The determination of $L$ is described later, leading to the query complexity estimate of Eq.~(\ref{eq:estQ}).

\begin{figure}[htbp]
\includegraphics[width=8.6cm]{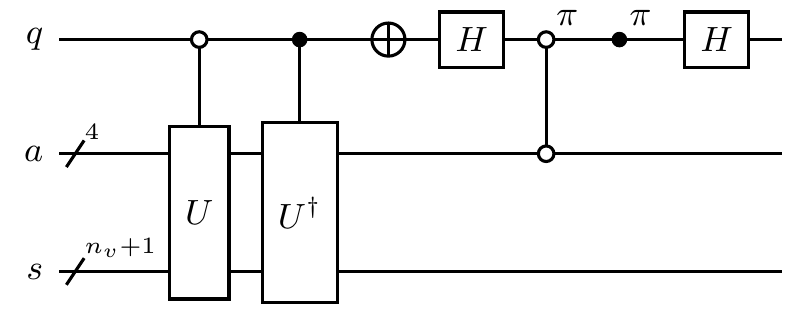}
\caption{Implementation of $W$. Gate $U := U^\dagger_{\text{row}} U_{\text{col}}$ where $U_{\text{row}}$ is given by Fig.~\ref{cir:Urow} and $U_{\text{col}}$ is given by Fig.~\ref{cir:Ucol}.}\label{cir:W}
\end{figure}

We define a query to be a single call to the operator $U$ or $U^\dagger$. The Hamiltonian is encoded in the top-left block of $U$:
\begin{equation}\label{eq:encoding}
\langle 0 \vert_a \langle j \vert_s U \vert 0 \rangle_a \vert k \rangle_s = \beta \hat{H}_{jk} := \hat{H}'_{jk}.
\end{equation}
The constant factor $\beta$ occurs because we are limited to encoding matrix rows and columns with normalizations of at most one. $\beta$ just amounts to a rescaling of the simulation time, which is corrected for by simulating for a time $t' := t/\beta$. To implement $U$ we use a strategy from \textcite{LowSpectral} of breaking it into components $U_{\text{row}}$ and $U_{\text{col}}$, each of which is a unitary state preparation operator:
\begin{align}
\begin{gathered}\label{eq:Us}
U := U^{\dagger}_{\text{row}} U_{\text{col}}, \\
U_{\text{row}} := \sum_{i, j} \vert \chi_{i,j} \rangle_{as} \langle i \vert_a \langle j \vert_s, \\
U_{\text{col}} := \sum_{i, k} \vert \psi_{i,k} \rangle_{as} \langle i \vert_a \langle k \vert_s.
\end{gathered}
\end{align}
The Hamiltonian that will be simulated is determined by the prepared states:
\begin{equation}
\hat{H}'_{jk} = \langle \chi_{0,j} \vert \psi_{0,k} \rangle.
\end{equation}

\begin{figure}[ht]
\includegraphics[width=8.6cm]{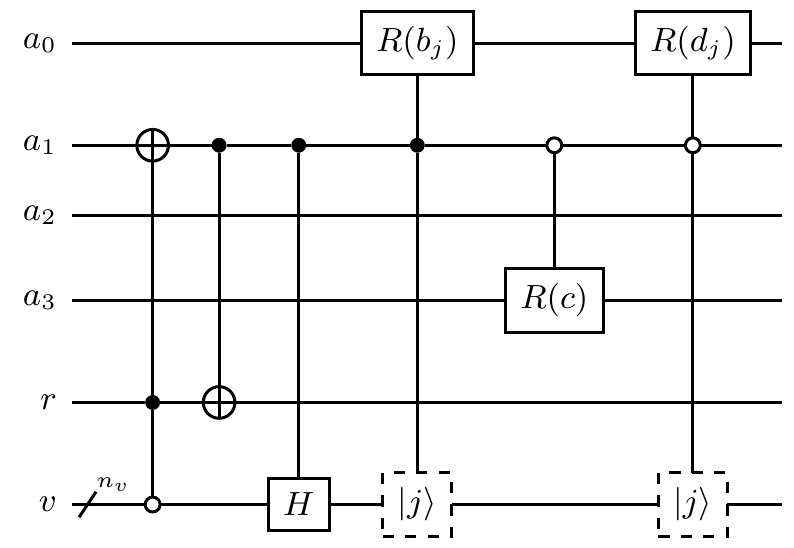}
\caption{Implementation of $\hat{U}_{\text{row}}$. The action of the rotation gate $R(\cdot)$ is given in Eq.~(\ref{eq:rotate}). Dashed boxes surround qubits that are inputs to a variable rotation, which does not modify those input qubits. The $a_i$ qubits are ancillas; only the behavior when they start as $\vert 0 \rangle$ is relevant.}\label{cir:Urow}
\end{figure}

\begin{figure}[ht]
\includegraphics[width=8.6cm]{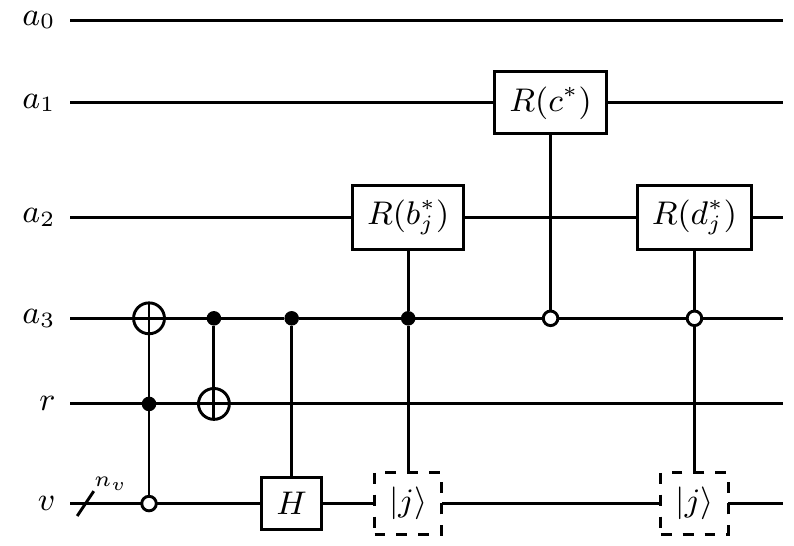}
\caption{Implementation of $\hat{U}_{\text{col}}$. The components are described in Fig.~\ref{cir:Urow}.}\label{cir:Ucol}
\end{figure}

There are still many possibilities for implementing the state preparation operators. Our choices are shown in Figs. \ref{cir:Urow} and \ref{cir:Ucol}. We introduce a rotation operator $R(\cdot)$, taken to be in SU(2), defined by its action:
\begin{equation}\label{eq:rotate}
R(\varrho) \vert 0 \rangle \rightarrow \varrho \vert 0 \rangle + \sqrt{1 - \vert \varrho\vert^2} \vert 1 \rangle
\end{equation}
for some $\vert \varrho\vert \leq 1$. When $\varrho$ is an efficiently-computable function of the qubits of a different register we call this a variable rotation. We only use values of $\varrho$ that are purely real or imaginary. This corresponds to applying $e^{-i \hat{\sigma}_y \arccos{\varrho}}$ or $e^{-i \hat{\sigma}_x \arccos{\text{Im}(\varrho)}} e^{i \hat{\sigma}_z \pi/2}$, respectively. By computing the rotation angles in temporary registers and applying rotations controlled on the angle qubits, these can be implemented efficiently. For our variable rotations, the input register has $n_v$ qubits, and the angles can be computed at a cost poly($n_v$) [assuming for simplicity that they are computed to poly($n_v$) bits of precision], so the variable rotations also cost $\text{poly}(n_v) = \text{polylog}(N_v)$.

The Hamiltonian implemented by $U$ takes the form
\begin{equation}
\hat{H}' = \sum_{j=0}^{N_v-1} \left[ \Omega_j \vert j \rangle \langle j \vert + \Upsilon_j \left(\vert j \rangle \langle N_v \vert + \vert N_v \rangle \langle j \vert \right) \right] + \hat{D}.
\end{equation}
Here $\hat{D} := \sum_{j=1}^{N_v-1} \Omega_j \vert j+N_v \rangle \langle j+N_v \vert$ is unimportant since it acts within the unused subspace of the state. We obtain our desired Hamiltonian [Eq.~(\ref{eq:Ham})] provided
\begin{equation}\label{eq:hp1}
\Omega_j = c^2 d_j^2 = k \beta v_j,
\end{equation}
\begin{equation}\label{eq:hp2}
\Upsilon_j = \sqrt{1 - \vert c \vert^2} \frac{d_j b_j}{\sqrt{N_v}} = \beta v_j \alpha_j,
\end{equation}
where $c$, $b_j$, and $d_j$ specify the rotations as depicted in Figs. \ref{cir:Urow} and \ref{cir:Ucol}. Since we also have the restriction $\vert b_j \vert, \vert d_j \vert < 1$, we choose
\begin{align}\label{eq:rotations}
d_j &= \sqrt{\frac{v_j}{v_{\text{max}}}}, & b_j &= \sqrt{\frac{v_j G_j}{g_{\text{max}}}},
\end{align}
where $v_{\text{max}} := \max_j \vert v_j\vert = (N_v-1)\Delta v/2$ and $g_{\text{max}} := \max_j \vert v_j G_j\vert$. Substituting Eq.~(\ref{eq:rotations}) into Eqs. (\ref{eq:hp1}) and (\ref{eq:hp2}) gives an equation for $c$:
\begin{equation}
c^2 = \text{sign}(k) \frac{\Gamma}{2} \left( \sqrt{1 + \frac{4}{\Gamma}} - 1\right),
\end{equation}
where $\Gamma := (k^2 v_{\text{max}}) / (\Delta v N_v g_{\text{max}})$. The last unknown, $\beta$, is given by $\beta = c^2/(k v_{\text{max}})$.

Since we must simulate $\hat{H}'$ for a time $t':=t/\beta$, the simulation costs scale with $1/\beta$. Consequently, the dependence of $\beta$ on problem parameters, including $k$, $N_v$, and $\Delta v$, is important. One can show\footnote{using $1 \leq (1+\sqrt{1/\Gamma})(\sqrt{1+4/\Gamma}-1)\Gamma/2 \leq 5/4 \; \forall{\Gamma \ge 0}$} that
\begin{equation}
\frac{4 \Lambda}{5} \leq \frac{1}{\beta} \leq \Lambda,
\end{equation}
where
\begin{equation}
\Lambda := \vert k \vert v_{\text{max}} + \sqrt{\Delta v N_v v_{\text{max}} g_{\text{max}}}.
\end{equation}
Crucially, since $\Delta v N_v = 2 v_{\text{max}} + \Delta v$, $\Lambda$ does not increase with decreasing $\Delta v$. Therefore using an extremely high-resolution grid does not increase the simulation query complexity. Meanwhile, the space and gate complexity scale only logarithmically with the grid resolution through the number of qubits $n_v$.

Since $\Vert \hat{H}' \Vert \leq 1$ is required by the Hamiltonian simulation technique, the ideal $1/\beta$ would be $\Vert \hat{H} \Vert$. An upper bound of $\Vert \hat{H} \Vert$, obtained by adding the norm of the diagonal part to the norm of the off-diagonal part, is
\begin{equation}
\Lambda' := \vert k \vert v_{\text{max}} + \sqrt{\sum_j v_j^2 G_j \Delta v}.
\end{equation}
Note that this expression turns into the $\Lambda$ expression if we replace $v_j^2 G_j$ with $v_{\text{max}} g_{\text{max}} = v_{\text{max}} \max_j \vert v_j G_j \vert$. So $\Lambda > \Lambda'$ but in most scenarios the difference is only a factor of order one. The variable rotations used to implement $U$ move some of the input state out of the good subspace, which leads to a smaller $\beta$. More general state preparation techniques (e.g. \cite{Soklakov2006, Grover2002}) could improve (increase) $\beta$, but for our purposes the simpler implementation choices are sufficient.

\section{Numerical Validation}\label{sec:tests}

\begin{figure}[htbp]
\includegraphics[width=8.6cm]{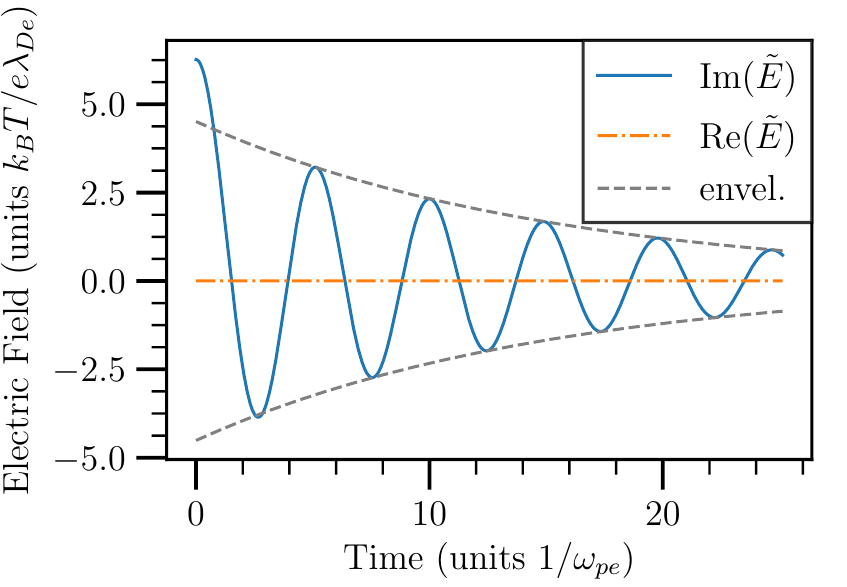}
\caption{Electric field evolution in the test case. $\tilde{E}$ is the Fourier component of the electric field for wavenumber $k$. The envelope (envel.) is a fitted exponential, which is accurate for later times. The vanishing of the real component of $\tilde{E}$ at all times is due to our specific choice of $\tilde{F}_j(t_0)$.}\label{fig:waveform}
\end{figure}

For testing, a code that implements the Hamiltonian simulation gates as linear algebra operations performed on a classical computer has been developed. This is compared to the result obtained by directly computing $e^{-i \hat{H} t}\vert x \rangle_s$ and to theoretical calculations. We test with the following parameter choices:
\begin{align*}
k &= 0.4, & v_{\text{max}} &= 4.5, & G_j,  \tilde{F}_j(t_0) &= \frac{1}{\sqrt{2\pi}} e^{-\frac12 v_j^2},\\
t &= 8\pi, & N_v &= 32, & \tilde{E}(t_0) = \frac{i}{k} &\sum_j \tilde{F}_j(t_0) \Delta v,
\end{align*}
where $t_0 = 0$ denotes the initial time, and the initial electric field is from Poisson's equation in 1D. The evolution of the electric field in this case is illustrated in Fig.~\ref{fig:waveform}.

\begin{figure}[htbp]
\includegraphics[width=8.6cm]{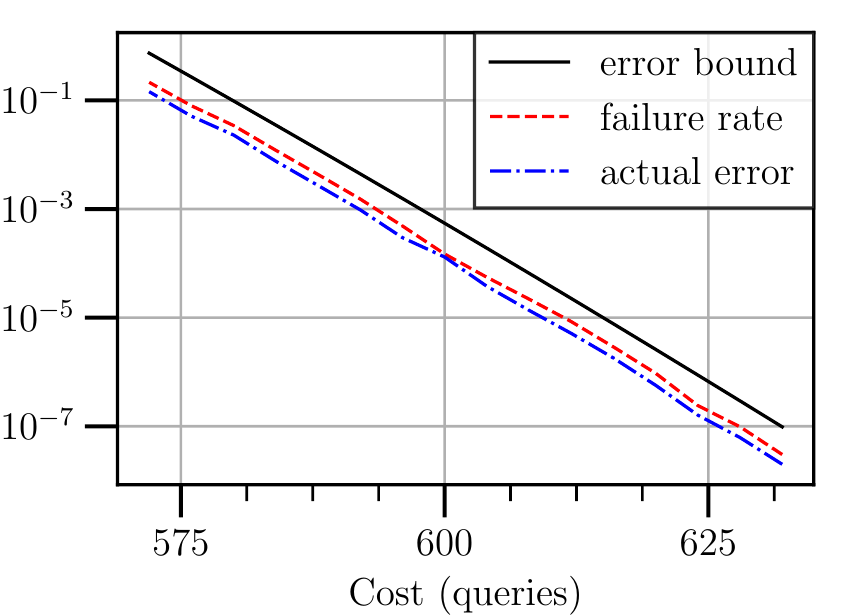}
\caption{Test of the simulation of the Hamiltonian in Eq.~(\ref{eq:Ham}). The error bound is given by Eq.~(\ref{eq:bound}), while the actual error $\varepsilon$ is computed using Eq.~(\ref{eq:sub_err}). The failure rate is $\vert (1 - \hat{P}_g )\hat{C}\vert x_g \rangle \vert^2 \leq 2\varepsilon$.}\label{fig:scaling}
\end{figure}

The quantum algorithm simulates the entire time period as one operation, accepting as parameters the total time and a maximally allowed error, with the error $\varepsilon$ for our circuit $\hat{C}$ defined to be
\begin{equation}\label{eq:sub_err}
\varepsilon := \left\vert \left(\hat{P}_g \hat{C} - e^{-i \hat{H} t}\right) \vert x_g \rangle \right\vert,
\end{equation}
where $\hat{P}_g$ is the projector onto the good subspace, specifically the space with all ancillas being zero, and $\vert x_g \rangle$ is any input within the good subspace, i.e. $\hat{P}_g \vert x_g \rangle$ = $\vert x_g \rangle$. For a given error tolerance $\epsilon$ the circuit can be designed to guarantee $\varepsilon \leq \epsilon$, based on the error bounds of the Hamiltonian simulation technique. We validate our implementation by running with a range of error tolerances. The results are displayed in Fig.~\ref{fig:scaling}.

The query complexity of our algorithm is $2L$ where $L = 2(n-1)$ and the positive integer $n$ is chosen such that the error bound\cite{Low2017},
\begin{equation}\label{eq:bound}
b(n) := 32 (t'/2)^n/n!,
\end{equation}
is below the chosen tolerance $\epsilon$. In our implementation we invert $b(n) \leq \epsilon$ numerically, but in the limit of $t' \gg 1$ a simple approximation is also effective. Note that $b(n) \leq e^{-\Delta n}b(n_0)$ with $n_0 := e t'/2+C$, $\Delta n := n - n_0 \ge 0$, and $C \ge 0$. This suggests taking $n = n_0 + \ln(1/\epsilon)$. Plugging in $\epsilon=1$ gives the bound $C \ge n - e(n!/32)^{1/n}$. The simple choice of $C=1.5$ satisfies this for all $n$. Then $b(n) < \epsilon$ for any $\epsilon \leq 1$ provided that
\begin{equation}\label{eq:estn}
n \ge e t' / 2 + \ln(1/\epsilon) + 1.5,
\end{equation}
which leads to a query complexity $Q$ with
\begin{equation}\label{eq:estQ}
Q \leq 2e t' + 4\ln(1/\epsilon) + 6.
\end{equation}
A somewhat stronger asymptotic bound, obtained by \textcite{Qubitization}, is
\begin{equation}
Q = \mathcal{O}\left(t' + \log(1/\epsilon)/\log\log(1/\epsilon)\right),
\end{equation}
but for $t' \gg 1$ and reasonable values of $\epsilon$ Eq.~(\ref{eq:estQ}) is fairly accurate, and the $t'$ term dominates. Our test case has $t' \approx 105.7$. The main features, shown in Fig.~\ref{fig:scaling}, of total query counts around 600 and error decreasing by approximately a factor of $e$ for every four additional queries, are consistent with Eq.~(\ref{eq:estQ}). All this does not depend on our specific Hamiltonian. However, as shown in Sec.~\ref{sec:Hsim}, the queries for our Hamiltonian can be implemented efficiently on a quantum computer, i.e. with costs scaling only as polylog($1/\Delta v$), while classically they cost $\mathcal{O}(1/\Delta v)$.

Having verified that the Hamiltonian simulation is accurate, now we interpret the results. Physically, the system has Langmuir oscillations that decay due to Landau damping. As Fig.~\ref{fig:waveform} shows, after a brief initial stage (i.e. the first period or so) the electric field is well described by a damped sinusoidal. Specifically, we can fit the function $A e^{-\gamma t}\cos(\omega t - \rho)$ to obtain parameters of interest, namely the frequency $\omega$ and damping rate $\gamma$:
\begin{align}
\omega &= 1.2851, & \gamma &= 0.0661.
\end{align}
Simple theoretical estimates\cite{Nicholson1983}, translated into our dimensionless variables and with $k=0.4$, give
\begin{align}
\omega \approx 1 + \frac32 k^2 &= 1.24, & \gamma \approx \sqrt{\frac{\pi}{8}} \frac{\omega}{k^3} e^{-\frac{\omega^2}{2k^2}} &= 0.099. \label{eq:gam}
\end{align}
These estimates rely on $k << 1$, so they can only give rough agreement. Precise values of $\gamma$ and $\omega$ can be found by numerical integration of the dispersion relation; in our units: $1 + \frac{1}{\sqrt{2\pi}k^2} \int_{-\infty}^{\infty} \frac{v \exp(-v^2/2) dv}{v - (\omega-i\gamma)/k} = 0$. This gives $\omega = 1.28506$ and $\gamma = 0.06613$, in agreement with the fit results. These comparisons show that our algorithm accurately reproduces Landau damping.

Note that this test case was chosen to be simple and easy to compute. The number of operations is not large, both for classical and quantum algorithms. Even the classical code that simulates the quantum computation completes in around a second. This is possible since we used a low-resolution velocity space, i.e. $N_v=32$. The classical costs would also be much higher for the multidimensional generalizations discussed in Sec.~\ref{sec:ext}.

\section{Initialization and Measurement}\label{sec:measure}
\begin{figure}[ht]
\includegraphics[width=8.6cm]{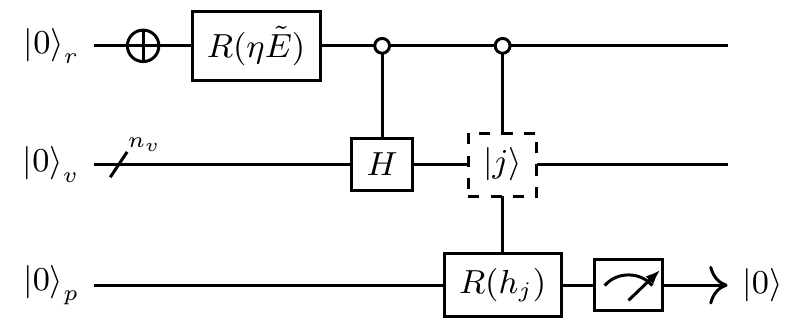}
\caption{Possible preparation circuit for the state in Eq.~(\ref{eq:init}). The rotation inputs $h_j$ are given in Eq.~(\ref{eq:rotH}). The operation is repeated until the measurement on the $p$ register returns $\vert 0\rangle$.}\label{cir:S}
\end{figure}

On a quantum computer, we would perform three steps: state preparation, Hamiltonian simulation, and measurement. State preparation costs depend on the prepared state, and in the worst case these costs scale with the state space size, but many classical probability distributions can be prepared efficiently\cite{Soklakov2006, Grover2002}. For simulating Landau damping we use an initially Maxwellian $\tilde{F}$, and the preparation can be done efficiently with Fig.~\ref{cir:S}. That circuit succeeds with probability $\langle \vert h_j \vert^2 \rangle$ where
\begin{equation}\label{eq:rotH}
h_j := \tilde{F}'_j / \max_i \vert \tilde{F}'_i \vert.
\end{equation}
Inputting $\tilde{F}'_j \propto \tilde{F}_j/\sqrt{G_j} \propto e^{v_j2/4}$ one finds $1/\langle \vert h_j \vert^2 \rangle = \mathcal{O}(v_{\max{}})$ so Fig.~\ref{cir:S}, which has a gate complexity $\text{poly}(n_v) = \text{polylog}(N_v)$, needs to be repeated $\mathcal{O}(v_{\max{}})$ times\footnote{or $\mathcal{O}(\sqrt{v_{\max{}}})$ rounds of amplitude amplification\cite{Brassard2002}.} before running the rest of the algorithm. Note that this cost just adds to the overall algorithm cost and does not scale with the simulation time. Consequently, it does not change the algorithm's asymptotic complexity.

Obtaining output from the algorithm is more difficult. In general quantum state tomography can be applied to learn the full state, but we do not want a scaling with the state space size. If we just want the electric field, that is a single amplitude, so the technique of amplitude estimation\cite{Brassard2002} is appropriate. This can produce an estimate $\tilde{p}$ of the probability $p$ of measuring the state to be within a specified subspace, with error bounded by
\begin{equation}\label{eq:est}
\delta := \vert \tilde{p} - p \vert \leq 2\pi \frac{\sqrt{p (1-p)}}{M} + \frac{\pi^2}{M^2},
\end{equation}
where $M$ is the number of iterations applied. Each iteration involves running the original algorithm forwards and backwards once, thus the full costs are multiplied by $2M$. From Eq.~(\ref{eq:est}), $M = \mathcal{O}(1/\delta)$, a quadratic speedup over direct sampling which would instead require $\mathcal{O}(1/\delta^2)$ repetitions to estimate the outcome probabilities with a standard deviation of $\sigma = \Theta(\delta)$.

If we want to efficiently estimate some probability to a specified \textit{relative} accuracy, we also need to ensure that it is not too small, since the costs scale with the absolute accuracy $\delta$. For our initially Maxwellian $\tilde{F}$, and with sums over velocity space approximated by integrals, one finds
\begin{equation}
\eta \vert \tilde{E}(t_0) \vert \approx \frac{1}{\sqrt{1+k^2}} \approx 0.928.
\end{equation}
Since this is not small, if we want to measure $\tilde{E}$ to a fixed accuracy $\delta'$ relative to the sinusoidal envelope, the measurement adds an $\mathcal{O}(1/\delta')$ cost factor provided the simulation time $t$ is chosen such that $e^{\gamma t} = \mathcal{O}(1)$, i.e. the electric field is only moderately damped.

\begin{figure}[ht]
\includegraphics[width=8.6cm]{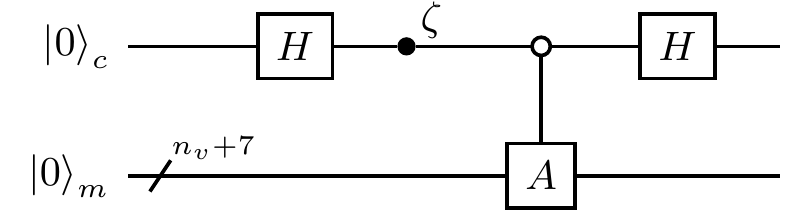}
\caption{Algorithm extension for obtaining complex phase information. Gate $A$ is the original algorithm, including state preparation but not measurement. The final state amplitude for all qubits equal to zero is modified by the phase $\zeta$ as given in Eq.~(\ref{eq:mph}).}\label{cir:M}
\end{figure}

Both direct sampling and amplitude estimation will only give information about magnitudes such as $\vert \tilde{E} \vert$. In general one may want information about the phase of a complex amplitude. This can be obtained by extending the original algorithm as shown in Fig.~\ref{cir:M}. We introduce another qubit, labeled by $c$, and a phase $\zeta$. If the original algorithm produced an amplitude $\nu$ for the $\vert 0 \rangle_m$ basis state, the extended algorithm modifies this via
\begin{equation}\label{eq:mph}
\nu\vert 0 \rangle_m \longrightarrow \frac{1}{2}\left(\nu + e^{i \zeta}\right) \vert 0 \rangle_c \vert 0 \rangle_m.
\end{equation}

Then the usual measurement process, e.g. amplitude estimation, can be applied to estimate $\left\vert\nu + e^{i \zeta}\right\vert$. By repeating this for $\zeta = \{0, \pm 2\pi/3\}$, we can determine the location of $\nu$ within the complex plane, rather than just its magnitude. Explicitly, if we write the new measured magnitudes as
\begin{align}
d_0 &:= \frac{1}{2}\left\vert\nu + 1\right\vert, & d_\pm &:= \frac{1}{2}\left\vert\nu + e^{\pm 2\pi i/3}\right\vert,
\end{align}
we can solve to obtain
\begin{equation}
\nu = \frac{2}{3} (2d_0^2 - d_+^2 - d_-^2) - \frac{2i}{\sqrt{3}} (d_+^2 - d_-^2).
\end{equation}
If each of the squared magnitudes is measured to an accuracy $\delta$, the same $\delta$ as in Eq.~(\ref{eq:est}), then we obtain an estimate $\tilde{\nu}$ of $\nu$ which obeys the bound $\vert\tilde{\nu} - \nu\vert \leq 8\delta/3$. Thus we can learn the value of $\nu$ within the complex plane to within a distance $\delta$ at a cost $\mathcal{O}(1/\delta)$. This can be applied to any chosen basis state, not just $\vert 0 \rangle_m$, by specifying the target state when performing amplitude estimation. For instance, we could use this technique to obtain the sign of Im($\tilde{E}$).

Now that all components of the algorithm have been discussed, we can determine its computational complexity, assuming the output is the final electric field. In principle the scaling with respect to any algorithm parameter can be considered, but some parameters are more relevant than others. When $k \gtrsim 1$ the electric field is rapidly damped away, so we assume $k = \mathcal{O}(1)$. Next, since the velocity components beyond $v_{\max{}}$ decay as $e^{-v_{\max{}}^2/2}$, a large $v_{\max{}}$ is generally unnecessary. In contrast, a large $N_v$ is needed to resolve the fine velocity-space oscillations that develop due to the $-i k v_j \tilde{F}'_j$ term in Eq.~(\ref{eq:fprime}). Therefore, we treat $v_{\max{}}$ as a constant and give the scaling with respect to $N_v$, the simulation time $t$, and the error tolerances $\epsilon$ and $\delta$. Taking $\epsilon = \delta$, the asymptotic gate complexity is
\begin{equation}
\mathcal{O}\left(\frac{\text{polylog}(N_v)}{\delta}\left(t + \frac{\log(1/\delta)}{\log\log(1/\delta)}\right)\right).
\end{equation}
If we assume that $t > \log(1/\delta)$, which is true for large enough $t$ and reasonable values of $\delta$, this simplifies to $\mathcal{O}(\text{polylog}(N_v) t/\delta)$.

The final electric field is just a simple example of an output for the algorithm. Given that $\tilde{E}$ oscillates in time, a more appropriate output may be
\begin{equation}\label{eq:rms}
\tilde{E}_{\text{rms}} := \sqrt{\sum_{i=0}^{M-1} \frac{\vert \tilde{E}(t_i) \vert^2}{M}},
\end{equation}
where the times $t_i$ are distributed over some time range of interest. In fact, this can be obtained by a simple extension of the original algorithm. Suppose each $t_i$ can be efficiently computed from its index $i$, and the times can be represented with integers $\tilde{t}_i := 2^m t_i/t_{\max} - 1$ where $t_{\text{max}} := \max_i t_i$. We can add an index register $i$ in a uniform superposition of the indices and a time register $t$, with $m$ qubits, in which the $t_i$ are computed. Then, controlled on each time qubit, the original algorithm is run for the time interval specified by that bit. The query bound of Eq.~(\ref{eq:estQ}) changes to $Q \leq 2e t_{\text{max}}' + m(4\ln(1/(m\epsilon))+6)$. After the Hamiltonian simulation, the state normalization within the subspace spanned by $\{\vert i \rangle_i \vert t_i \rangle_t \vert N_v \rangle_s\} \forall{i}$ and all ancilla qubits $\vert 0 \rangle$ will be $\eta \tilde{E}_{\text{rms}}$ (to within error $\epsilon$). Amplitude estimation can be applied to determine this normalization with a better scaling than direct sampling. Thus we can get $\tilde{E}_{\text{rms}}$ with the same dominant cost scaling as the original algorithm, i.e. $\mathcal{O}(\text{polylog}(N_v)t_{\text{max}}/\delta)$.

\section{Generalizations}\label{sec:ext}
In Sec.~\ref{sec:formulation} we initially worked in 3D velocity space, with a uniform ${\bf B}_0$ and general field perturbations ${\bf E}_1$ and ${\bf B}_1$. In principle that problem can still be solved efficiently on a quantum computer using Hamiltonian simulation. A difficulty is that the dropped terms are large: $c_n \gg 1$ for nonrelativistic plasmas and $\Vert \frac{\partial}{\partial {\bf v}} \Vert = \mathcal{O}(1/\Delta v)$. Consequently, the spectral norm of the Hamiltonian would be dramatically increased by including these terms, and Hamiltonian simulation costs scale with that norm. However, there is a potential solution: we can break the Hamiltonian into the new large piece $A$ and the old small piece $B$ and apply the Hamiltonian simulation technique of \textcite{Low2018}. In that paper, the interaction picture is applied along with time-dependent Hamiltonian simulation to allow separate handling of the two components. We would take $A$ to be the Hamiltonian in $d \vert x \rangle / dt = -i A \vert x \rangle$ where the variables are encoded in $\vert x \rangle$ and are evolving via
\begin{align}
\frac{\partial \tilde{\bf B}}{\partial t} &= -i c_n {\bf k} \times \tilde{\bf E}, & \frac{\partial \tilde{\bf E}}{\partial t} &= i c_n {\bf k} \times \tilde{\bf B},
\end{align}
\begin{equation}\label{eq:gyro}
\frac{\partial \tilde{f}^{\prime}}{\partial t} = \frac{1}{c_n} ({\bf v} \times {\bf B}_0) \cdot \frac{\partial \tilde{f}'}{\partial {\bf v}},
\end{equation}
while $B$ would be the Hamiltonian for the other terms of Eqs. (\ref{eq:Vlasov3}) and (\ref{eq:Amp2}).

As is, our algorithm handles evolution by $B$ in the 1D case. The extension of the velocity space to 3D is simple: the 1D velocity index is replaced by a flattened 3D velocity index. The matrix would have $\mathcal{O}(1/\Delta^3v)$ nonzero entries in one row and column, but the coupling constants of Eq.~(\ref{eq:coupling}) scale as $\sqrt{\Delta^3v}$ so we still have $\Vert B \Vert = \mathcal{O}(1)$. Then we can obtain similar efficiency, i.e. costs linear in time and logarithmic in error and grid resolution, provided that we can efficiently implement $e^{i A \tau}$ with $\tau = \mathcal{O}(1)$, corresponding to the time evolution by the added terms alone.

The EM evolution of a single Fourier mode is given by Eq.~(\ref{eq:EM}). This is trivial to solve; the pairs $\{E_y, B_x\}$ and $\{E_x, B_y\}$ each undergo rotations about the x-axis in their Bloch spaces. Therefore rotation gates can be applied to perform this evolution, circumventing the usual scaling with spectral norm. We still need a way to efficiently perform the evolution of Eq.~(\ref{eq:velop}) (which commutes with the EM evolution). We can reexpress this using
\begin{equation}\label{eq:Lv}
\frac{i}{c_n} ({\bf v} \times {\bf B}_0) \cdot \frac{\partial}{\partial {\bf v}} = \frac{1}{c_n} {\bf B}_0 \cdot {\bf \hat{L}_v},
\end{equation}
where ${\bf \hat{L}_v} := \frac{{\bf v}}{i} \times \frac{\partial}{\partial {\bf v}}$ is the velocity space analog of the angular momentum operator. Thus the evolution caused by this term is simply the rotation of velocity space around ${\bf B}_0$, i.e. it generates cyclotron motion. Once again we can hope to implement this efficiently in a direct way, avoiding spectral norm scaling. For instance, if we work on a cylindrical velocity grid, symmetric about the ${\bf B}_0$ axis and with velocity cells spaced uniformly in $\phi$, then certain discrete time steps just correspond to addition of a binary constant to the qubits associated with the $\phi$ index. We leave the details for potential future work.

In general the distribution function is 6D. Our algorithm applies to a single spatial Fourier mode ${\bf k}$, but it can easily be extended to many modes. For applying the algorithm to a superposition of modes, the only adjustment is that all quantities which depend on ${\bf k}$ need to be computed in superposition using operations controlled on the new ${\bf k}$ register(s). In principle this can still be done efficiently, but we will not work this out here. Additionally, if the initial variables are specified as functions of position, we could use the quantum Fourier transform\cite{NielsenChuang} to efficiently convert this to a Fourier space representation before running the algorithm. Another consideration is the magnitude of ${\bf k}$. If we do not assume $\vert {\bf k} \vert \lesssim 1$ then the simulation costs go up since the Hamiltonian spectral norm scales linearly with $\vert {\bf k} \vert $.  Physically, $\vert {\bf k} \vert \lesssim 1$ is justified in that this is enough to resolve $\lambda_{De}$. Moreover, when $\vert {\bf k} \vert \gtrsim 1$ the electrostatic waves are rapidly damped away, as suggested by Eq.~(\ref{eq:gam}). Then even with many modes, having $\vert {\bf k} \vert \lesssim 1 \;\forall{\bf k}$ is reasonable.

Another significant restriction made in Sec.~\ref{sec:formulation} was the assumption of a Maxwellian background distribution. This makes the system unconditionally stable. The alternative, where some growing wave modes exist, violates unitarity since then a state that holds the perturbed quantities must grow exponentially if it contains a growing mode component. Such cases cannot be handled directly with Hamiltonian simulation. A completely different algorithm, possibly utilizing a quantum linear systems algorithm, would be required, and it is unclear whether it would scale well.  This is a topic of future work. Still, our algorithm could be applied to other stable background distributions. The background distribution enters Eqs. (\ref{eq:Faraday}), (\ref{eq:Vlasov3}), and (\ref{eq:Amp2}) only through its velocity derivative. In our dimensionless variables, $\frac{\partial f_M}{\partial {\bf v}} = - {\bf v} f_M$. For a more general background distribution $f_0$ we can make the replacement $f_M \rightarrow g_0$ where
\begin{equation}
\frac{\partial f_0({\bf v})}{\partial {\bf v}} = -{\bf v} g_0({\bf v}).
\end{equation}
By redoing the steps up through Eq.~(\ref{eq:coupling}) one finds that the new Hamiltonian is Hermitian provided $g_0({\bf v}) \ge 0 \; \forall{\bf v}$. For background distributions that satisfy this restriction, Hamiltonian simulation can still be applied. Furthermore, in the electrostatic case and with ${\bf B}_0 = {\bf 0}$ the background distribution enters only through $\frac{\partial f_0}{\partial v_z}$. Here the appropriate replacement is
\begin{equation}
G_{j_z} \longrightarrow \frac{-1}{\partial v_z}\sum_{j_x, j_y} f_0({\bf v_j}) \Delta^2 v,
\end{equation}
and the unitarity restriction is $G_{j_z} \ge 0 \; \forall{j_z}$. Our algorithm can be easily applied to such distributions; only the specific values of $b_j$ and $\beta$ would be altered.

Thus far we have solved for the motion of electrons, with ions assumed stationary. If we remove this restriction then each species obeys its own version of Eq.~(\ref{eq:fullVlasov}), and they are coupled through the fields. Keeping the same dimensionless variables, Eq.~(\ref{eq:Vlasov2}) for each species is
\begin{equation}
\frac{\partial \tilde{f}_s}{\partial t} = -i {\bf k} \cdot {\bf v} \tilde{f}_s + r_s \left(\tilde{\bf E} \cdot {\bf v} f_M - \frac{1}{c_n} ({\bf v} \times {\bf B}_0) \cdot \frac{\partial \tilde{f}_s}{\partial {\bf v}}\right),
\end{equation}
where $r_s := \frac{m_e q_s}{m_s e}$. If we absorb the $r_s$ factor into the background distribution and magnetic field, we recover the original mathematical form, and the prior manipulations are still applicable, including the discussed extensions. The rescaling factors $\mu(\bf v)$ become species dependent through the $r_s$ factor. One may be concerned about the sign change for positive species, given the importance of signs for unitarity, but since there is a corresponding sign change in Eq.~(\ref{eq:Amp}), there are no new issues here. The generators of the evolution of each $\tilde{f}_s$, including the couplings with $\tilde{E}$, are anti-Hermitian, therefore the combined evolution of all species and fields is unitary. Additionally, the terms in the evolution, and thus the matrix entries, for other species are relatively small since $r_s \ll 1$ for ions. Consequently the Hamiltonian spectral norm, along with the space and time complexity of the algorithm, would be nearly unchanged by the inclusion of multiple species.

\section{Summary}\label{sec:sum}

In high-temperature plasma physics, evolution of the particle distribution function is governed by the Vlasov-Maxwell system of equations, but simulating this classically with the full 6D phase space is extremely expensive. For instance, if the phase-space grid requires 1000 cells in each dimension, just storing $f({\bf x},{\bf v}, t)$ is beyond exascale capability\cite{Service2018}. However, on a quantum computer, the distribution function can be encoded as the amplitudes of a quantum state, requiring exponentially fewer qubits than classical bits. Moreover, we have shown that the linearized Vlasov-Maxwell system of equations (with $f_0=f_M$) produces a unitary evolution of the field variables and a rescaled version of the distribution function. Thus Hamiltonian simulation algorithms can be applied to perform this classical physics simulation.

For one limiting form of the full system we demonstrated in detail how the Hamiltonian simulation can be performed efficiently, such that the quantum algorithm gets an exponential speedup in space and time complexity over the classical simulation with respect to the velocity grid size. To obtain this efficiency we used a recently developed Hamiltonian simulation technique, and the efficient handling of the more general problem appears to require even more sophisticated Hamiltonian simulation techniques. An additional challenge is the extraction of an output. For measuring a value encoded in a single quantum amplitude to within absolute error $\delta$, $\mathcal{O}(1/\delta)$ iterations of the original algorithm are required. This can significantly diminish the quantum speedup, depending on the desired accuracy.

\begin{acknowledgments}
Research supported in part by the United States Department of Energy under grant DE-SC0020393.
\end{acknowledgments}

\bibliography{bib1}

\end{document}